%% file: Charm-proceedings_v6.tex
\documentclass[12pt]{article}
\usepackage{graphicx}

\def\pbnr{}
\def\speaker{Elaine C. F. S. Fortes$^1$, K.S. Babu$^2$, and R. N. Mohapatra$^3$}
\def\onbehalfof{}
\def\title{Flavor Physics Constraints on TeV Scale Color Sextet Scalars\\[0.05in]

\vspace{0.3cm}}
\def\affiliation{$^1$Instituto de F\'isica Te\'orica-Universidade Estadual Paulista, R. Dr. Bento Teobaldo Ferraz 271, S\~ao Paulo-SP, 01140-070, Brazil\\ $^2$Department of Physics, Oklahoma State University,
Stillwater, OK 74078, USA\\$^3$ Maryland Center for Fundamental Physics and Department of Physics\\ University of Maryland, College Park, MD 20742, USA}
\def\support{}

\input charmmacros.tex

\begin{document}
\begin{titlepage}
\pubblock

\vfill
\Title{\title}
\vfill
\Author{\speaker\SupportedBy{\support}\OnBehalf{\onbehalfof}}
\Address{\affiliation}
\vfill
\begin{Abstract}
In this work we present an analysis of flavor violating effects mediated by color
sextet scalars, which arise naturally in left-right symmetric gauge
theories based on $SU(2)_L \times SU(2)_R \times SU(4)_C$
group. The sextets, denoted here by $\Delta_{dd}$, $\Delta_{ud}$ and $\Delta_{uu}$, couple to right--handed quarks.
We delineate the constraints on these couplings arising from
meson--anti-meson transitions and flavor changing weak decays. The sextet $\Delta_{uu}$  mediates $D^{0}-\overline{D}^{0}$ mixing via tree-level
and box diagrams, and also mediates $D \rightarrow K \pi,\, \pi\pi$ decays. The sextets $\Delta_{ud}$ and $\Delta_{dd}$ mediate $B_{d}^{0}-\overline{B}_{d}^{0}$, $B_{s}^{0}-\overline{B}_{s}^{0}$ and $K^{0}-\overline{K}^{0}$ mixings as well as rare $B$ and $D$ meson
decays.
Our analysis shows that for coupling strengths of order $10^{-2}$, the current experimental data requires the mass of these color sextets to exceed several TeV. These bounds are stronger than those obtained from direct LHC searches.
\end{Abstract}
\vfill
\begin{Presented}
\venue
\end{Presented}
\vfill
\end{titlepage}
\def\thefootnote{\fnsymbol{footnote}}
\setcounter{footnote}{0}
%

\section{Introduction}
${}$
${}$
The high energy physics community awaits with great anticipation the discovery of new physics at the LHC. Additional heavy scalar particles are among potential candidates to be discovered. In this work we consider a class of color--sextet scalars that arise in natural extensions  of the standard model to accommodate small neutrinos masses, and study the constraints on their masses and couplings to quarks. If these scalars have masses in the TeV range they would lead to flavor changing neutral current effects (FCNC) through
tree and one--loop processes. Since many FCNC observables have been measured or are constrained with high precision, they imply strong constraints on the couplings of the color sextets. They in turn will have implications for models that use them to explain different phenomena such as origin of matter in the universe or top anti-top forward--backward asymmetry.

The main motivation for the existence of color sextet particles comes from the seesaw mechanism for generating neutrino masses.
When the Standard Model (SM) is embedded into the $SU(2)_{L}\times SU(2)_{R}\times SU(4)_{c}$ framework \cite{ps}, where parity symmetry and quark--lepton
symmetry are manifest, new scalars transforming as (1,3,10) are needed to break the gauge symmetry and to generate Majorana masses for the
right--handed neutrinos. This scalar multiplet contains color sextet diquark fields $\Delta_{qq}$, along with leptoquark and SM singlet fields.
FCNC processes mediated by these diquarks is the focus of this analysis.
Such color sextet diquarks are among the basic ingredients of a new way of understanding the origin of matter generated after sphaleron decoupling, called post-sphaleron baryogenesis~\cite{Nasri1}.  Getting adequate amount of matter requires certain ranges of the color sextet masses and couplings.  It is then important to know if the desired parameters are compatible with FCNC constraints. In fact, a recent work presented in Ref. \cite{Babu2013} shows that  the constraints of post-sphaleron baryogenesis, when combined with the neutrino oscillation data and with the restrictions from FCNC mediated by color sextet scalars imply an upper limit on the baryon number violating process of neutron--antineutron oscillation  \cite{marshak}, which may be accessible to the next generation of proposed experiments.

Color sextet scalars can also be searched for at the LHC and their signature at colliders are presented in Ref. \cite{Mohapatra0}. Again the signal strength for these fields would depend on their masses and couplings. Further work on the collider production has been carried out in Ref.\,\cite{Zhan,Zhan2}, where there have been estimates of the next-to-leading order QCD corrections and the threshold resummation effects at LHC. Reference \cite{Zhan2} presents results from the recent dijet data from  LHC to give constraints on the couplings between colored scalars and quarks. This gives us enough motivations to study the phenomenological implications of the color sextet scalars for low energy physics.

\section{Color Sextet Couplings and Induced FCNC}

There are three different kinds of color sextet fields present in the (1,3,10) multiplet, each coupling to right--handed quarks: $uu$,
$dd$ and $ud$ (denoted as
$\Delta_{uu},\Delta_{dd},\Delta_{ud}$ respectively). These scalars mediate FCNC both at the tree--level and
at one--loop level, which probe different flavor combinations of couplings.
For instance, since  the $\Delta_{dd}$ field couples to $dd$, $ss$ and $bb$ quarks, it will mediate
$B_{s,d}^{0}-\overline{B}_{s,d}^{0}$ and $K^{0}-\overline{K}^{0}$
mixings as well as $B$ meson decays. The couplings of the $\Delta_{ud}$ field will
be constrained  by the box diagram contributions to $B_{s,d}^{0}-\overline{B}_{s,d}^{0}$ and
$K^{0}-\overline{K}^{0}$ mixings as well as by $B$ and $D$ meson decays and
also by $K\rightarrow \pi\pi $ decays.  The couplings of
$\Delta_{uu}$ are constrained by $D^{0}-\overline{D}^{0}$
mixing and decays such as $D\rightarrow\pi\pi, K\pi$.

In the context of $SU(2)_{L}\times SU(2)_{R}\times SU(4)_{c}$ model,
in order to be consistent with the limits from FCNC processes
and  $n-\overline{n}$ oscillations, in Ref. \cite{Nasri1,Babu2013} it is suggested that two of the three scalars should have mass
of order TeV while the third one (e.g. $\Delta_{uu}$) should have a mass in the order of 100 TeV.
Here we have performed our analysis taking the masses of the three scalars to be of order TeV.
Clearly, these bounds on the couplings can be easily scaled when masses are different.

The Yukawa couplings of the right--handed quarks with the color sextets $\Delta_{(dd,uu,ud)}$ can be written as:
\begin{eqnarray}\label{1b}
   \mathcal{L}_{\Delta}=&&-\frac{f_{ij}}{2}(d)_{Ri}^{T} C (d)_{Rj} \Delta_{dd}~-\frac{h_{ij}}{2}(u)_{Ri}^{T} C (u)_{Rj} \Delta_{uu}~
  -\frac{g_{ij}}{2\sqrt{2}}(u_{Ri}^{T} C d_{Rj} \cr && +\, d_{Ri}^{T} C u_{Rj}) \Delta_{ud}\texttt{ + h.c.,}
\end{eqnarray}
where $f_{ij}$ and $h_{ij}$  are
symmetric in the flavor indices $(i,j)$,  $C$ is the charge-conjugation operator, and
$(d,u)_{R}=(1+\gamma_{5})/2 $$(d,u)$.  In the $SU(2)_{L}\times SU(2)_{R}\times SU(4)_{c}$ limit we have
$f_{ij} = h_{ij} = g_{ij}$.

The effective $\Delta F=2$ Hamiltonian resulting from integrating out the $\Delta_{dd}$ field can be written as
$\mathcal{H}_{\Delta F=2}=\mathcal{H}_{tree}+\mathcal{H}_{box}$ where
\begin{eqnarray}\label{1d}
&&\mathcal{H}_{\Delta
F=2}=-\frac{1}{8}\frac{f_{i\ell}f_{kj}^{*}}{M_{\Delta_{dd}}^{2}}(\overline{d}_{kR}^{\alpha}\gamma_{\mu}d_{iR}^{\alpha})(\overline{d}_{j
R}^{\beta}\gamma^{\mu}d_{\ell R}^{\beta})
 +\frac{1}{256\pi^{2}}\frac{[(ff^{\dagger})_{ij}(ff^{\dagger})_{\ell k}+(ff^{\dagger})_{ik}(ff^{\dagger})_{\ell j}]}{M_{\Delta_{dd}}^{2}}
\cr&& \times\{(\overline{d}_{jR}^{\alpha}\gamma_{\mu}d_{iR}^{\alpha})(\overline{d}_{kR}^{\beta}\gamma^{\mu}d_{\ell R}^{\beta})+ 5(\overline{d}_{jR}^{\alpha}\gamma_{\mu}d_{iR}^{\beta})(\overline{d}_{kR}^{\beta}\gamma^{\mu}d_{\ell R}^{\alpha})\}~.
 \end{eqnarray}
 For applications to $\Delta F=2$ ($F=B$, S) processes mediated by $\Delta_{dd}$, we set the flavor indices to be $i=\ell=3$ and $j=k=2$ for $B_{s}-\overline{B}_{s}$ mixing, $i=\ell=3$ and $j=k=1$ for $B_{d}-\overline{B}_{d}$ mixing, and $i=\ell=2$ and $j=k=1$ for $K-\overline{K}$ mixing
 in Eq. (\ref{1d}). The effective Hamiltonian for the $\Delta_{uu}$ field which
 mediates $\Delta C=2$ processes can be read from
Eq. \ref{1d}  with the substitution of $d$-type quarks
by $u$-type quarks and the replacement of $f$ by
$h$. For $D-\overline{D}$ mixing we set $i=\ell=1$ and $j=k=2$.

The effective Hamiltonian resulting from the exchange of $\Delta_{ud}$ can be written as:
\begin{eqnarray}\label{1g}
    \mathcal{H}_{eff}&=&-\frac{1}{32}\frac{\widehat{g}_{ij}\widehat{g}_{kl}^{*}}{M_{\Delta_{ud}}^{2}}\left[(\overline{u}_{kR}^{\alpha}\gamma_{\mu}u_{iR}^{\alpha})(\overline{d}_{\ell R}^{\beta}\gamma^{\mu}d_{j R}^{\beta})+(\overline{u}_{kR}^{\alpha}\gamma_{\mu}d_{iR}^{\alpha})(\overline{d}_{\ell R}^{\beta}\gamma^{\mu}u_{j R}^{\beta})\right]\nonumber \\
    && +\frac{1}{256\pi^{2}} \frac{1}{64}\frac{1}{M_{\Delta_{ud}}^{2}}\times[(\widehat{g}\widehat{g}^{\dagger})_{ij}(\widehat{g}\widehat{g}^{\dagger})_{\ell k}+(\widehat{g}\widehat{g}^{\dagger})_{ik}(\widehat{g}\widehat{g}^{\dagger})_{\ell
    j}]\times\nonumber\\
   &&\left[(\overline{d}_{jR}^{\alpha}\gamma_{\mu}d_{iR}^{\alpha})(\overline{d}_{kR}^{\beta}\gamma^{\mu}d_{\ell R}^{\beta})+5(\overline{d}_{jR}^{\alpha}\gamma_{\mu}d_{iR}^{\beta})(\overline{d}_{kR}^{\beta}\gamma^{\mu}d_{\ell
   R}^{\alpha})\right]
  \end{eqnarray}
where $\widehat{g}_{ij}=(g_{ij}+g_{ji})/2$. New contributions to neutral meson mixings  will be  generated by the second part of this Hamiltonian.
We set $i=\ell=3$ and $j=k=2$ for $B_{s}^{0}-\overline{B}_{s}^{0}$
mixing, $i=\ell=3$ and $j=k=1$ for
$B_{d}^{0}-\overline{B}_{d}^{0}$ mixing, and $i=\ell=1$ and $j=k=2$
for $K^{0}-\overline{K}^{0}$ mixing in Eq. (\ref{1g}).

\section{Bounds from Neutral Meson Mixing and Decays}
In order to derive the bounds on the couplings and masses of the color sextet scalars, we performed a Fierz transformation in the Hamiltonians given in Eqs. (\ref{1d}) and (\ref{1g}), and applied the relation $\Delta m_{X}=2|\left\langle\overline{X}|H_{eff}^{\Delta X=2}|X\right\rangle|$ for $X-\overline{X}$ mass difference, where $X$ labels $B_{s}^{0},B_{d}^{0},K^{0},D^{0}$. The numerical values of the masses, the decay constants and the bag parameters for meson mixings are taken from Ref. \cite{Becirevic}. The calculated constraints are presented in Table~\ref{tab-mix}.

\begin{table}[h!]
\begin{center}
\begin{tabular}{|c|c|c|} \hline\hline
Process & Diagram & Constraint on Couplings \\ \hline
 & Tree &  $|f_{22} f^*_{33} |\leq 7.04\times 10^{-4}\left(\frac{	M_{\Delta_{dd}}}{1~{\rm TeV}}\right)^2$\\
$\Delta m_{B_s}$ & Box & $\sum^3_{i=1}|f_{i3} f^*_{i2}|\leq 0.14 \left(\frac{M_{\Delta_{dd}}}{1~{\rm TeV}}\right)$\\
& Box & $\sum^3_{i=1}|\hat{g}_{i3} \hat{g}^*_{i2}|\leq 1.09 \left(\frac{M_{\Delta_{ud}}}{1~{\rm TeV}}\right)$ \\ \hline
& Tree & $ |f_{11} f^*_{33}|\leq 2.75\times 10^{-5}\left(\frac{	M_{\Delta_{dd}}}{1~{\rm TeV}}\right)^2$\\
$\Delta m_{B_d}$ & Box & $\sum^3_{i=1}|f_{i3} f^*_{i1}|\leq 0.03 \left(\frac{M_{\Delta_{dd}}}{1~{\rm TeV}}\right)$\\
& Box & $\sum^3_{i=1}|\hat{g}_{i3} \hat{g}^*_{i1}|\leq 0.21 \left(\frac{M_{\Delta_{ud}}}{1~{\rm TeV}}\right)$\\ \hline
& Tree & $ |f_{11} f^*_{22}|\leq 6.56\times 10^{-6}\left(\frac{M_{\Delta_{dd}}}{1~{\rm TeV}}\right)^2$\\
$\Delta m_{K}$ & Box & $\sum^3_{i=1}|f_{i2} f^*_{i1}|\leq 0.01 \left(\frac{M_{\Delta_{dd}}}{1~{\rm TeV}}\right)$\\
& Box & $\sum^3_{i=1}|\hat{g}_{i1} \hat{g}^*_{i2}|\leq 0.10 \left(\frac{M_{\Delta_{ud}}}{1~{\rm TeV}}\right)$\\ \hline
$\Delta m_{D}$ & Tree &  $|h_{11}h_{22}^*| \leq 3.72\times 10^{-6}\left(\frac{M_{\Delta_{uu}}}{1~{\rm TeV}}\right)^2$\\
& Box & $\sum^3_{i=1} |h_{i2}h_{i1}^*|\leq 0.01\left(\frac{M_{\Delta_{uu}}}{1~{\rm TeV}}\right)$ \\ \hline\hline
\end{tabular}
\end{center}
\caption{Constraints on the product of Yukawa couplings of the color sextet scalars arsing from $K^0-\overline{K}^0,~D^0-\overline{D}^0,~B^0_s-\overline{B}^0_s$ and $B^0_d-\overline{B}^0_d$ mixing.} \label{tab-mix}
\end{table}

The effective Hamiltonian of Eqs. (\ref{1d})- (\ref{1g}) can also induce flavor changing nonleptonic decay at tree level. After reproducing the Standard Model results using the factorization approach ~\cite{Ali1}, we evaluated  the matrix elements at momentum scale $\mu$. For the $V+A$ operators we adopted the convention presented in Ref.~\cite{Jang} to express the matrix elements of quark bilinear operator in terms of meson decay constants and form factors. Our results are presented in Tables \ref{tab-bdecay} and \ref{tab-ddecayud}. We note that the sextets $\Delta_{dd}$ and $\Delta_{ud}$ also contribute to the pure annihilation processes $\overline{B}_{s}^{0}\rightarrow \pi^{0}\pi^{0}$ and $\overline{B}_{s}^{0}\rightarrow \pi^{+}\pi^{-}$. But, considering the interference between SM and new physics, we conclude that these decays won't impose stringent limits on the couplings of the these sextets when compared to other decays presented here. In a forthcoming paper we will present further details of the analysis for all color sextet scalars.

\begin{table}[t]
\begin{center}
\begin{tabular}{|c|c|c|}\hline\hline
Sextet & Decay  & constraints on couplings \\ \hline
 $\Delta_{dd}$ & $B^-\to \pi^0\pi^-$ & $|f_{13}f^*_{11}|\leq 0.705$  \\
&$\overline{B}^0_d\to \phi\pi^0 $  &$|f_{23}f^*_{12}|\leq 0.035$ \\
&$B^-\to \phi\pi^-$ & $|f_{23}f^*_{12}|\leq 0.030$  \\
&$\overline{B}^0_d\to \phi\overline{K}^0 $  &$|f_{23}f^*_{22}|\leq 0.294$ \\
&$B^-\to \phi K^-$ & $|f_{23}f^*_{22}|\leq 0.308$  \\
&$\overline{B}^0_d\to \pi^0\pi^0 $  &$|f_{13}f^*_{11}|\leq 0.444$ \\
&$\overline{B}^0_d\to \overline{K}^0 K^0 $  &$|f_{23}f^*_{12}|\leq 0.279$ \\
&$\overline{B}^0_d\to K^0K^0 $  &$|f_{13}f^*_{22}|\leq 0.559$ \\
&$B^-\to K^0 K^-$ & $|f_{23}f^*_{12}|\leq 0.271$  \\
&$B^-\to \overline{K}^0 K^-$ & $|f_{13}f^*_{22}|\leq 0.541$  \\
&$\overline{B}^0_d\to \overline{K}^0\pi^0 $  &$|f_{13}f^*_{12}|\leq 0.313$ \\
&$B^-\to \pi^0K^-$ & $|f_{13}f^*_{12}|\leq 0.461$  \\
&$B^-\to \pi^-\overline{K}^0$ & $|f_{13}f^*_{12}|\leq 1.268$  \\ \hline
 $\Delta_{ud}$ &$B^{-}\rightarrow \pi^{-}\pi^{0}$ & $|(g_{13}+g_{31})g_{11}^{*}|<1.058$\\
  &$\overline{B}_{d}^{0}\rightarrow \pi^{-}\pi^{+}$ & $|(g_{13}+g_{31})g_{11}^{*}|<1.456$\\
  &$\overline{B}_{d}^{0}\rightarrow \pi^{0}\pi^{0}$ & $|(g_{13}+g_{31})g_{11}^{*}|<1.331$\\
  &$\overline{B}_{d}^{0}\rightarrow \overline{K}^{0}\pi^{0}$ &$|(g_{13}+g_{31})(g_{12}^{*}+g_{21}^{*})|<5.068$  \\
  &$\overline{B}_{d}^{0}\rightarrow {K}^{-}\pi^{+}$ & $|(g_{13}+_{31})(g_{12}^{*}+g_{21}^{*})|<4.794$ \\
&${B}^{-}\rightarrow {K}^{-}\pi^{0}$ &$|(g_{13}+_{31})(g_{12}^{*}+g_{21}^{*})|<2.711$\\
  & $\overline{B}_{s}^{0}\rightarrow {K}^{-}K^{+}$ & $|(g_{13}+g_{31})(g_{12}^{*}+g_{21}^{*})|<7.097$ \\\hline\hline
\end{tabular}\label{tab-bdecay}
\end{center}
\caption{Constraints on the product of the $f$ and $g$-couplings from non-leptonic $B$-meson decays.  These constraints are
obtained in the QCD factorization method. The numbers in the second column should be multiplied by a factor $(M_{\Delta_{dd,ud}}/{\rm TeV})^2$.}
\label{tab-bdecay}
\end{table}

\begin{table}[h]
\begin{center}
\begin{tabular}{|c|c|c|}\hline\hline
Sextet & Decay  & constraints on couplings \\ \hline
$\Delta_{ud}$ &$D^{0}$$\rightarrow$ $\overline{K}^{0}\pi^{0}$&$\mid g_{12}+g_{21}\mid^{2} <134.005$\\
  & $D^{0}$$\rightarrow$$ K^{-}\pi^{+}$&  $\mid g_{12}+g_{21}\mid^{2}<274.715$ \\
   &$D^{+}$$\rightarrow$$ \overline{K}^{0}\pi^{+}$ &  $\mid g_{12}+g_{21}\mid^{2}<50.485$ \\
   &$D^{+}$$\rightarrow$$ K^{0}\pi^{+}$&  $\mid g_{22}g_{11}\mid <92.927$\\
   &$D^{+}$$\rightarrow$$ K^{+}\pi^{0}$ &$\mid g_{22}g_{11}\mid <11.237$\\
   &$D^{0}$$\rightarrow$$ K^{+}\pi^{-}$&  $\mid g_{22}g_{11}\mid <14.497$\\
   &$D^{0}$$\rightarrow$$ K^{0}\pi^{0}$& $\mid g_{22}g_{11}\mid <126.268$\\
   &$D^{+}$$\rightarrow$$ \pi^{+}\pi^{0}$& $|(g_{12}+g_{21})g_{11}^{*}|< 143.548$\\
   & $D^{0}$$\rightarrow$$ \pi^{+}\pi^{-}$& $|(g_{12}+g_{21})g_{11}^{*}| < 108.036$\\
   &$D^{0}$$\rightarrow$$ \pi^{0}\pi^{0}$& $|(g_{12}+g_{21})g_{11}^{*}| < 84.164$\\
   & $D^{0}$$\rightarrow$$ K^{+}K^{-}$ & $|(g_{12}+g_{21})g_{22}^{*}| < 147.957$ \\
   & $D^{+}_{s}$$\rightarrow$$ K^{+}\phi$ & $|(g_{12}+g_{21})g_{22}^{*}| < 16.347$\\\hline
    $\Delta_{uu}$  & $D^{+}$$\rightarrow$$ \pi^{+}\pi^{0}$& $|h_{12}h_{11}^{*}| < 47.849$\\
    &$D^{0}$$\rightarrow$$ \pi^{0}\pi^{0}$& $|h_{12}h_{11}^{*}| < 28.054$\\
    & $D^{+}_{s}$$\rightarrow$$ K^{+}\pi^{0}$& $|h_{12}h_{11}^{*}| < 33.485 $\\\hline \hline
\end{tabular}\label{tab-ddecayud}
\end{center}
\caption{Constraints on the product of the $g$ and $h$-couplings from non-leptonic  $D$-meson decays, obtained
in the QCD factorization method. The numbers in the second column should be multiplied by a factor $(M_{\Delta_{ud,uu}}/{\rm TeV})^2$.}
\label{tab-ddecayud}
\end{table}

\section{Conclusion}
We have summarized in Tables \ref{tab-mix}--\ref{tab-ddecayud} the results of a detailed study of the constraints  on the TeV mass color sextet couplings to right--handed quarks coming from tree and box diagram contributions to the decays of mesons containing $b$ and $c$-quarks as well as $\overline{B}_{d}^{0}-B_{d}^{0}$, $\overline{B}_{s}^{0}-B_{s}^{0}$, $\overline{D}^{0}-D^{0}$ and $\overline{K}^{0}-K^0$ transitions using current experimental data.  The constraints coming from the meson mixings and from $B$-meson decays are more stringent when compared to the ones which come from $D$-decays.

\Acknowledgements
The work of KSB is supported in part by the US Department of Energy Grant No. DE-FG02-04ER41306, RNM is supported
by National Science Foundation grant No. PHY-0968854. The work of ECFSF is supported by FAPESP under contract No. 2011/21945-8.

\end{document}

%% file: charmmacros.tex
\newcommand\pubnumber{\pbnr}
\newcommand\pubdate{\today}
\def\Title#1{\begin{center} {\Large #1 } \end{center}}
\def\Author#1{\begin{center}{ \sc #1} \end{center}}

\newcommand{\OnBehalf}[1]{\sbox0{#1}\ifdim\wd0=0pt
        {}
	\else
	{\\on behalf of #1}
	\fi}
\newcommand{\SupportedBy}[1]{\sbox0{#1}\ifdim\wd0=0pt
        {}
	\else
	{\footnote{#1}}
	\fi}
\def\Address#1{\begin{center}{ \it #1} \end{center}}

\newcommand\pubblock{\includegraphics[width=5cm]{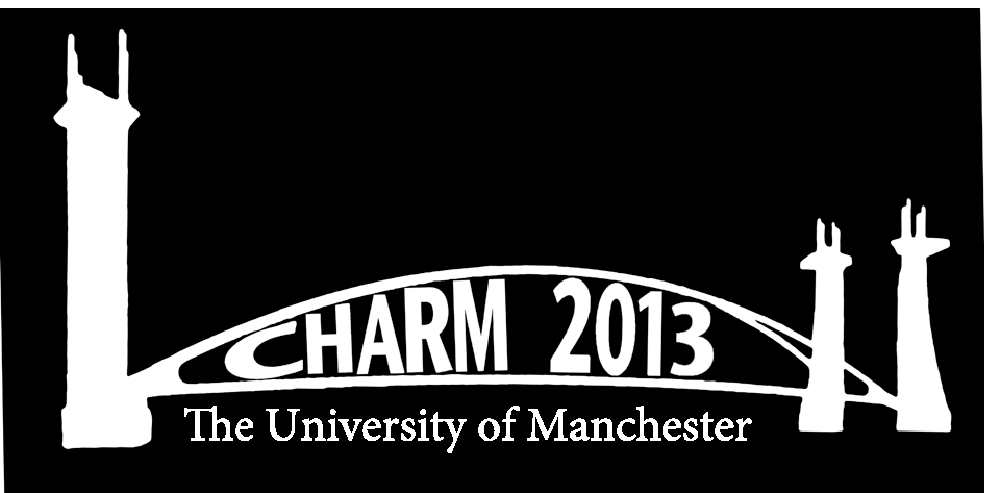}\hfill{\begin{tabular}{l} \pubnumber\\
         \pubdate  \end{tabular}}}
\newenvironment{Abstract}{\begin{quotation}  }{\end{quotation}}
\newenvironment{Presented}{\begin{quotation} \begin{center} 
             PRESENTED AT\end{center}\bigskip 
      \begin{center}\begin{large}}{\end{large}\end{center} \end{quotation}}
\def\Acknowledgements{\bigskip  \bigskip \begin{center} \begin{large}
             \bf ACKNOWLEDGEMENTS \end{large}\end{center}}
\def\venue{The 6$^{th}$ International Workshop on Charm Physics\\
(CHARM 2013)\\
Manchester, UK,  31 August -- 4 September, 2013}

\def\beq{\begin{equation}}
\def\eeq#1{\label{#1}\end{equation}}
\def\eeqn{\end{equation}}

\def\beqa{\begin{eqnarray}}
\def\eeqa#1{\label{#1}\end{eqnarray}}
\def\eeqan{\end{eqnarray}}

\let\bar=\overbar

\def\Dslash{\not{\hbox{\kern-4pt $D$}}}
\def\dslash{\not{\hbox{\kern-2pt $\del$}}}

\def\msb{{\bar{\ssstyle M \kern -1pt S}}}

